# Predictions of pressure-induced transition temperature increase for a variety of high temperature superconductors

Mario Rabinowitz[a] and T. McMullen[b]

[a]*Electric Power Research Institute, Palo Alto, CA 94303 USA lrainbow@stanford.edu*

[b]*Department of Physics, Virginia Commonwealth University, Richmond, Virginia 23284*

## Abstract

A wide variety of superconducting oxides are used to test a general model of high pressure induced transition temperature ($T_c$) changes. The $T_c$'s vary from a low of 24 K to a high of 164 K. Although the model is capable of predicting both increases and decreases in $T_c$ with pressure, only superconductors that exhibit an increase are considered at this time. Predictions are made of the maximum $T_{cP}^{theo}$ for 15 superconductors as a function of their compressibilities. The theoretical results generally agree well with experiment. This model of $T_c$ as a function of pressure is derived from a recent successful phenomenological theory of short coherence length superconductivity.

## 1. Introduction

The dependence of the superconducting transition temperature on pressure, $T_c(P)$, is an important functional dependence that is well documented experimentally, but has not been well understood theoretically for high temperature superconductors (HTSC). In an excellent in-depth review article on this field, Schilling and Klotz [1] observe, "Unfortunately, only a few theoreticians have been courageous enough to predict $T_c(P)$ for HTSC; we hope more will follow." We accept this challenge, and note that this dearth of theories for $T_c(P)$ follows from the scarcity of theories which can predict $T_c$ at ordinary pressures for HTSC.

Most HTSC are hole-doped. Griessen [2], Wijngaarden and Griessen [3], Murayama et al [4], and Mori [5] have all observed and experimentally verified that generally $T_c$ increases with pressure for hole-doped HTSC. (There are some exceptions as with Tl-Ba-Cu-O, and some stoichiometries of Y-Ba-Cu-O.) Furthermore, the logarithmic derivative $d \ln T_c / dP$ is usually inversely correlated with the value of $T_c$, increasing as $T_c$ decreases, for both hole-doped and electron-doped HTSC. For electron-doped HTSC, Markert et al [6] and others [3] have noted that $T_c$ ordinarily decreases with the application of pressure. Exceptions to these rules can arise due to structural transitions, inter- and intra-planar spacings, and other causes. A non-negligible amount of compression takes place just upon cooling -- particularly for HTSC -- the volume contraction for HTSC on cooling is 2 - 3 times larger than for the common metals. [1]

The application of hydrostatic pressure permits a potentially uncluttered variation of the state of a superconductor without the complications of changing two or more parameters at the same time that chemical substitution may entail. This allows a testing of theoretical models for which in fortuitous cases only one parameter is varied. We feel that in these cases, it is an excellent test of the phenomenological theory of short coherence length superconductivity developed by Rabinowitz that successfully yields $T_c$ for a wide variety of HTSC such as cuprates, bismates, and thallates; as well as a large number of heavy fermion, organic, chevral, and dichalcogenide low $T_c$ superconductors. [7-12] It also works well for superfluids such as $^3$He and $^4$He. [9,13]

**2. Analysis**

The transition temperature proposed by Rabinowitz [7-9] for short coherence length materials is

$$T_c = A\left[\frac{\hbar^2 n^{2/3}}{2mk}\right], \tag{1}$$

where $A = A_3 = 0.218$ for isotropic three-dimensional (3-D) superconductors and $A_2 = (3/2)A_3 = 0.328$ for anisotropic 2-D superconductors, $\hbar$ is (Planck's constant)/$2\pi$, m is the carrier effective mass, and k is the Boltzmann constant. Although the functional dependence of the variables is testable in this analysis, the specific value of A cancels out. The number density of carriers is $n = N/V$, where N is the number of carriers and V is the volume of the sample.

In order to encompass a wide variety of HTSC, let us consider a more general derivation than we previously presented.[14] Equation (1) implies

$$\frac{T_{cP}}{T_c} = \left(\frac{N_P}{N_o}\right)^{2/3}\left(\frac{m_o}{m_P}\right)\left(\frac{V_o}{V_P}\right)^{2/3} . \tag{2}$$

The inverse of the bulk modulus is the compressibility

$$\kappa = -\frac{1}{V}\frac{\partial V}{\partial P} . \tag{3}$$

Integrating eq. (3), we obtain

$$\left(\frac{V_o}{V_P}\right) = e^{\kappa' \Delta P}, \tag{4}$$

where κ' is the average compressibility over the range ΔP from 0 to P. Substituting eq. (4) in eq. (2)

$$\left(\frac{T_{cP}}{T_c}\right) = e^{\frac{2}{3}\kappa' \Delta P}\left(\frac{N_P}{N_o}\right)^{2/3}\left(\frac{m_o}{m_P}\right). \tag{5}$$

Equation (5) is in the most general form resulting from eq. (1). Let us now make an approximation so that we may readily calculate the change in $T_c$ with pressure using just the compressibility. In so far as $\left(\frac{N_P}{N_o}\right), \left(\frac{m_P}{m_o}\right), \left(\frac{\kappa'}{\kappa}\right)$ are $\approx 1$, then

$$T_{cP}^{theo} \approx T_c^{exp} e^{\frac{2}{3}\kappa\Delta P} \approx T_c^{exp}(1 + \tfrac{2}{3}\kappa\Delta P), \qquad (6)$$

where $T_{cP}^{theo}$ is the maximum transition temperature attainable by applying the pressure differential $\Delta P$, and $T_c^{exp}$ is the experimental transition temperature at atmospheric pressure. The right hand side approximation to the exponential yields eq. (4) in reference 14.

### 3. Discussion

Here we shall use the exponential form in eq. (6) for calculating $T_{cP}^{theo}$ in determining the extent to which the compressibility alone contributes to the high pressure increase of $T_c$ to its maximum value. The results are shown in Table 1 where the theoretically derived values are compared with the experimental values. The $\Delta P$ column lists the pressure in GPa (gigaPascals: 1GPA = 10 kbar = $10^9$ n/m$^2$) used to attain the maximum $T_c$. Table 1 shows excellent agreement between $T_{cP}^{theo}$ and $T_{cP}^{exp}$ for #1,2,3,4,6, and 7. Our predicted changes in $T_c$ for the other compounds are smaller than those observed. This could be due to a variety of causes. Among these could be (in the context of our eq. 5) changes in the carrier density beyond the simple volume effect included here; or changes in the effective mass with pressure. Other possibilities outside the scope of our theory could be pressure-induced structural transitions, or significant changes in the interaction strength. These need to be explored.

A discussion of the sources for the data used to compile Table 1 should prove helpful to the reader. The excellent review article, *The Influence of High Pressure on the Superconducting and Normal State Properties of High Temperature Superconductors*, by Schilling and Klotz [1] has served both as a guide for finding relevant data, as well as helping to interpret this data. Most of the superconductors examined here, and their properties may be found in reference 1, except for the more recent data on the Hg-Ba-Ca-Cu-O system.

The results with the Hg-Ba-Ca-Cu-O system present a unique challenge for explanation. The TCSUH group [15] have taken the optimally doped

$HgBa_2Ca_{q-1}Cu_qO_{2q+2+\delta}$ (where q = 1, 2, and 3) system up to ≈ 164K for Hg-1223, ≈ 154K for Hg-1212, and ≈ 118K for Hg-1201. This is the main source of data for #1-3, except for the compressibility κ. From the auspicious Cornelius and Schilling model [16], we have estimates of the compressibility κ for $HgBa_2Ca_{q-1}Cu_qO_{2q+2+\delta}$ of (1/88) $(GPa)^{-1}$ , (1/94) $(GPa)^{-1}$, and (1/101) $(GPa)^{-1}$ with q = 1, 2, and 3 respectively.

$Bi_{1.68}Pb_{0.32}Ca_{1.85}Sr_{1.75}Cu_{2.65}O_{10}$, #4, is in Fig. 12 of [1] and in Mori et al [17] where it is listed in Table 3 as Bi2223. Cornelius, Klotz, and Schilling [18] in their very thorough tabulation of their calculated compressibilities compared with experimental values list $Bi_{1.9}Pb_{0.3}Ca_{1.9}Sr_2Cu_3O_{10.25}$ which is the closest we could find.

Numbers 5 - 15 may be found in reference 1, Tables 1, 2, and 3, and Figs. 12 and 14; as well as in references 4, and 19-26. Compressibilities are supplemented in reference 18.

## 4. Conclusion

We have shown that by using the Rabinowitz model of short coherence length superconductivity[7-14], it is possible to predict the high pressure maximum transition temperature, $T_{cP}^{theo}$, with reasonable accuracy using only the compressibility for 14 out of 15 superconductors. For the 15th superconductor, $T_{cP}^{theo}$ was found within a factor of 2 of the experimental value. For # 1 -14, the theoretical values deviated from the experimental values from less than 1% to less than 25%. Inclusion of the additional variables may bring the agreement even closer to the experimental values. These 15 superconductors are not a select list of favorable cases, but rather all the superconductors for which we could readily obtain data. We would appreciate receiving data on the maximum $T_c$, ΔP, and κ that readers may be able to bring to our attention, and invite their comments.

Of course other effects such as pressure-induced suppression of undesirable phases, or production of favorable phases may dominate in some systems. Thus far we have only looked at materials for which pressure increases the transition temperature.

It should not be surprising that pressure alone is capable of doing this since the 16 elements Si, P, S, Ca, Sc, Ge, As, Se, Sr, Y, Sb, Te, Cs, Ba, Bi, and Lu as well as more complex materials become superconducting with the application of pressure. In addition to structural phase transformations, which may lead to improved superconducting properties, the increased number density of charge carriers appears to play a prominent role in increasing $T_c$.

**Acknowledgement**

We wish to express our sincere gratitude to James S. Schilling for generously and expeditiously providing reprints of his work on the effects of high pressure on superconductors. These have proved invaluable in the writing of our paper.

Table 1

Comparison of experimental and theoretical maximum pressure-induced transition temperatures[a)]

| Compound | $T_c^{exp}$ (K) | $T_{cP}^{exp}$ (K) | $T_{cP}^{theo}$ (K) | $\kappa\ 10^{-3}$ (GPa)$^{-1}$ | $\Delta P$ (GPa) |
|---|---|---|---|---|---|
| 1. $HgBa_2Ca_2Cu_3O_{8+\delta}$ | 135 | 164 | 166 | 10 | 31 |
| 2. $HgBa_2CaCa_2O_{8+\delta}$ | 128 | 154 | 157 | 10.6 | 29 |
| 3. $HgBa_2CuO_{8+\delta}$ | 94 | 118 | 113 | 11.4 | 24 |
| 4. $Bi_{1.68}Pb_{0.32}Ca_{1.85}Sr_{1.75}Cu_{2.65}O_{10}$ | 111 | 119 | 120 | 14.7 | 8 |
| 5. $Tl_2Ba_2CaCu_2O_8$ | 109 | 119 | 111 | 7.3 | 3 |
| 6. $YBa_2Cu_3O_{7-\delta}$ | 92 | 95 | 94.1 | 6.7 | 5 |
| 7. $YBa_2Cu_3O_{7-\delta}$ | 90 | 93 | 93.7 | 6.7 | 9 |
| 8. $Y_{0.9}Ca_{0.1}Ba_2Cu_4O_8$ | 90 | 98 | 92.7 | ≈ 9 | 5 |
| 9. $YBa_2Cu_4O_8$ | 80 | 101 | 83.2 | 8.5 | 7 |
| 10. $YBa_2Cu_4O_8$ | 80 | 106 | 84.7 | 8.5 | 10 |
| 11. $CaBaLaCu_3O_{6.85}$ | 62 | 86 | 64.8 | 8.2 | 8 |
| 12. $La_{1.85}Sr_{0.15}CuO_4$ | 36 | 41 | 36.8 | 6.9 | 4.5 |
| 13. $La_{1.85}Sr_{0.15}CuO_4$ | 36 | 41 | 36.7 | 6.79 | 4 |
| 14. $La_{1.8}Ba_{0.2}CuO_4$ | 36 | 44 | 36.1 | ≈ 5.3 | 1.1 |
| 15. $Nd_{1.32}Sr_{0.41}Ce_{0.27}CuO_{3.96}$ | 24 | 46 | >24.9 | 7.01 | >8 |

[a)] This is a compilation of all the superconductors for which sufficient data could be obtained to calculate the maximum pressure-induced transition temperature $T_{cP}^{theo}$ from the experimental atmospheric pressure transition temperature $T_c^{exp}$, in the pressure excursion $\Delta P$, using only the compressibility $\kappa$. The calculated values $T_{cP}^{theo}$ are compared with the experimental values $T_{cP}^{exp}$.

Comparison of experimental and theoretical maximum pressure-induced transition temperatures[a)]

| Compound | $T_c^{exp}$ (K) | $T_{cP}^{exp}$ (K) | $T_{cP}^{theo}$ (K) | % | κ $10^{-3}$ $(GPa)^{-1}$ | ΔP (GPa) |
|---|---|---|---|---|---|---|
| 1. $HgBa_2Ca_2Cu_3O_{8+\delta}$ | 135 | 164 | 166 | +1.2 | 10 | 31 |
| 2. $HgBa_2CaCa_2O_{8+\delta}$ | 128 | 154 | 157 | +2.1 | 10.6 | 29 |
| 3. $HgBa_2CuO_{8+\delta}$ | 94 | 118 | 113 | -4.5 | 11.4 | 24 |
| 4. $Bi_{1.68}Pb_{0.32}Ca_{1.85}Sr_{1.75}Cu_{2.65}O_{10}$ | 111 | 119 | 120 | +1 | 14.7 | 8 |
| 5. $Tl_2Ba_2CaCu_2O_8$ | 109 | 119 | 111 | -7.2 | 7.3 | 3 |
| 6. $YBa_2Cu_3O_{7-\delta}$ | 92 | 95 | 94.1 | -1 | 6.7 | 5 |
| 7. $YBa_2Cu_3O_{7-\delta}$ | 90 | 93 | 93.7 | +1 | 6.7 | 9 |
| 8. $Y_{0.9}Ca_{0.1}Ba_2Cu_4O_8$ | 90 | 98 | 92.7 | -5.4 | ≈ 9 | 5 |
| 9. $YBa_2Cu_4O_8$ | 80 | 101 | 83.2 | -17.6 | 8.5 | 7 |
| 10. $YBa_2Cu_4O_8$ | 80 | 106 | 84.7 | -20.1 | 8.5 | 10 |
| 11. $CaBaLaCu_3O_{6.85}$ | 62 | 86 | 64.8 | -24.7 | 8.2 | 8 |
| 12. $La_{1.85}Sr_{0.15}CuO_4$ | 36 | 41 | 36.8 | -10.4 | 6.9 | 4.5 |
| 13. $La_{1.85}Sr_{0.15}CuO_4$ | 36 | 41 | 36.7 | -10.6 | 6.79 | 4 |
| 14. $La_{1.8}Ba_{0.2}CuO_4$ | 36 | 44 | 36.1 | -18.0 | ≈ 5.3 | 1.1 |
| 15. $Nd_{1.32}Sr_{0.41}Ce_{0.27}CuO_{3.96}$ | 24 | 46 | >24.9 | -45.8 | 7.01 | >8 |

a) This is a compilation of all the superconductors for which sufficient data could be obtained to calculate the maximum pressure-induced transition temperature $T_{cP}^{theo}$ from the experimental atmospheric pressure transition temperature $T_c^{exp}$, in the pressure excursion ΔP, using only the compressibility κ . The calculated values $T_{cP}^{theo}$ are compared with the experimental values $T_{cP}^{exp}$, and the percentage deviation, %, is given.